\documentstyle[12pt]{article}
\renewcommand \baselinestretch{1.4}

\begin{document}

\def\beq{\begin{equation}}
\def\eeq{\end{equation}}
\def\bce{\begin{center}}
\def\ece{\end{center}}
\def\bea{\begin{eqnarray}}
\def\eea{\end{eqnarray}}
\def\ben{\begin{enumerate}}
\def\een{\end{enumerate}}
\def\ul{\underline}
\def\ni{\noindent}
\def\nn{\nonumber}
\def\bs{\bigskip}
\def\ms{\medskip}
\def\wt{\widetilde}
\def\wh{\widehat}
\def\tr{\mbox{Tr}\, }
\def\brr{\begin{array}}
\def\err{\end{array}}





\vspace*{3mm}

\begin{center}

{\LARGE \bf
 Dilatonic gravity near two dimensions and asymptotic freedom of
the gravitational coupling constant}

\vspace{8mm}

\renewcommand
\baselinestretch{0.5}
\medskip
{\sc E. Elizalde}
\footnote{E-mail: eli@zeta.ecm.ub.es} \\
Center for Advanced Study CEAB, CSIC, Cam\'{\i} de Santa
B\`arbara,
17300 Blanes
\\ and Department ECM and IFAE, Faculty of Physics,
University of Barcelona, \\ Diagonal 647, 08028 Barcelona,
Catalonia, Spain \\
 and \\
{\sc S.D. Odintsov} \footnote{E-mail: odintsov@ecm.ub.es} \\
Tomsk Pedagogical Institute, 634041 Tomsk, Russia
\\
and Department ECM, Faculty of Physics,
University of  Barcelona, \\  Diagonal 647, 08028 Barcelona,
Catalonia,
Spain \\

\vspace{15mm}

{\bf Abstract}

\end{center}

Two models of dilatonic gravity are investigated: (i)
dilaton-Yang-Mills gravity and (ii) higher-derivative dilatonic
gravity.
Both are renormalizable in $2+\epsilon$ dimensions and have a
smooth
limit for $\epsilon \rightarrow 0$. The corresponding one-loop
effective
actions and  beta-functions are found. Both theories are shown to
possess a non-trivial ultraviolet fixed point ---for all dilatonic
couplings--- in which the gravitational constant is asymptotically
free.
It is shown that in the regime of asymptotic freedom  the matter
central
charge can be significantly increased by two different mechanisms
---as
compared with pure dilatonic gravity, where $n < 24$.

\vspace{4mm}


\newpage

\ni{\bf Introduction}.
 It seems clear by now that one of the ways to provide a somewhat
natural solution to the problem of the non-renormalizability of
Einstein
gravity is to be found in the $2+\epsilon$ approach. This follows
mainly
from the well-known observation \cite{3} that the gravitational
coupling
constant $G$ in $2+\epsilon$ dimensions exhibits an asymptotically
free
behavior. Unfortunately, it has been
shown that Einstein gravity in
$2+\epsilon$  dimensions  (a theory that has no smooth $\epsilon
\rightarrow 0$ limit) is not renormalizable near two
dimensions
\cite{13}. This renders the usefulness of asymptotic freedom of
Einstein's gravity in $2+\epsilon$ dimensions a very doubtful
issue.

Recently, it has been suggested  \cite{9} that an interesting thing
to
do could be to study dilatonic gravity (which indeed has a smooth
$\epsilon \rightarrow 0$ limit) near two dimensions. (Similarly, it
was the idea in Ref. \cite{14} to consider Einstein gravity with
a scalar field near two dimensions).
As has been shown \cite{9},
dilatonic gravity ---which is a renormalizable theory for
$d=2+\epsilon$--- possesses a non-trivial,
ultraviolet stable fixed point in the domain where it is
asymptotically
free on the gravitational coupling constant. The only drawback of
these
theories is that in the asymptotically free (AF) regime the matter
central charge (that is, the central charge of our universe) is
restricted to be $n <24$ \cite{9} or $n<25$ \cite{14}. It is
therefore
important to search for theories which, having similar good
properties
as those, can overcome the problem of such strongly rigid
restrictions
on the matter central charge.

In the present leter we will consider two models of dilatonic
gravity:
(i) a dilaton-Yang-Mills (DYM) theory and (ii) higher-derivative
(HD)
dilatonic gravity
 in $2+\epsilon$ dimensions. We will show that both these theories
have
properties very similar to those of pure dilatonic gravity, in
particular, they are multiplicatively renormalizable and have a
smooth
limit when $\epsilon \rightarrow 0$. Moreover, they also have
corresponding non-trivial, ultraviolet stable fixed points, where
the
gravitational coupling constant is asymptotically free and the
dilatonic
couplings are specified. The number of scalars in both models can
be
significantly increased, in the case of the DYM theory by
enlarging the number of gauge fields, and in HD dilatonic gravity
by a
convenient choice of the HD dilatonic functions.
\bs

\ni{\bf Dilatonic-Yang-Mills gravity near two dimensions}.
The action we are going to consider, in $2+\epsilon$ dimensions, is
 \beq
S = \int d^d x\,  \sqrt{-g} \left[{\mu^\epsilon \over 16\pi G} R
e^{-2\phi}
- \frac{1}{2} g^{\mu\nu}
\partial_\mu \chi_i \partial_\nu \chi^i e^{-2\Phi (\phi)} + {1
\over 4}
 e^{-f_2(\phi)} (G^a_{\mu\nu})^2 \right], \label{1}
\eeq
where $\mu$ is a mass parameter, $g_{\mu\nu}$ is the
($2+\epsilon$)-dimensional metric,
$R$ the corresponding curvature, $\chi_i$ are scalars ($i=1,2,
\ldots, n$), $G^a_{\mu\nu} = \nabla_\mu A^a_\nu - \nabla_\nu
A^a_\mu +
f_{bc}^a  A^b_\mu  A^c_\nu $, with
$A_\mu^a$ a gauge field, ($i=1,2,\ldots, N$), the gauge group being
supposed to be simple and compact, with antisymmetric structure
constants
 $f_{bc}^a  $.
The smooth functions $\Phi (\phi)$ and $f_2(\phi)$  are the
scalar-dilaton and vector-dilaton couplings, respectively.
This action corresponds, in fact, to renormalizable
DYM
gravity in  $2+\epsilon$ dimensions. It is easy to see that the
most general
 form, which would
include a dilaton kinetic term multiplied by an arbitrary
dilatonic function
can be actually converted in  $2+\epsilon$ dimensions to the form
(\ref{1}). To show this, the only thing one has to do is a local
Weyl
tranformation of the metric.  In addition, a
 gravity-dilaton interacting term $f_1 (\phi )R$ can be
always
represented as in  (\ref{1}) by a suitable redefinition of the
dilaton.
Notice also that, unlike pure Einstenian gravity \cite{3}, the
theory
(\ref{1}) has a smooth $\epsilon \rightarrow 0$ limit. Hence, the
calculation of the one-loop effective action in exactly-two
dimensions
gives also the divergences of the effective action near two
dimensions.

The calculation of the one-loop divergences in pure dilatonic
gravity
has been initiated in Ref. \cite{8} and discussed later in Refs.
\cite{2,5}, for pure dilatonic gravity, and in Refs. \cite{4,6,7},
for dilatonic gravity interacting via a non-trivial dilatonic
coupling with other fields.

We can use the results obtained in Refs. \cite{4,6} in order to
write the
divergent part of the effective action in the theory under
discussion.
When employing the background-field method
\beq
g_{\mu\nu} \longrightarrow  g_{\mu\nu} +  h_{\mu\nu} ,
\ \ \ \ \ \  \phi  \longrightarrow  \phi + \varphi
 \label{2} \eeq
where $ h_{\mu\nu}$ and $\varphi$ are quantum fields, it is
convenient
to use, respectively, the following gauge-fixing actions
in the gravitational
and gauge field sectors:
\bea
S_{gf}^1 & =& - {\mu^\epsilon \over 32\pi G} \int d^d x\,
\sqrt{-g} \,
 g_{\mu\nu} \chi^\mu \chi^\nu e^{-2\phi}, \nn \\
 \chi^\nu  & =& \nabla_\mu \bar{h}^{\mu\nu} + 2 \nabla^\nu \varphi,
\nn
\\ S_{gf}^2 & =&  \int d^d x\,  \sqrt{-g} \, (\nabla_\mu Q^{a\mu}
)^2
e^{-f_2(\phi )}, \label{3}
\eea
where $\bar{h}^{\mu\nu}$ is the traceless part of $h^{\mu\nu}$ and
$ Q^{a\mu}$ is a quantum gauge field.

Using the gauge fixing actions (\ref{3}), the one-loop effective
action
can be calculated following Refs. \cite{4,6} (specifically, in Ref.
\cite{6}
the explanation is given about how to extend the results already
obtained for
dilatonic-Maxwell gravity to the present case of a
DYM one).

The final result for the one-loop effective action reads as follows
\bea
\Gamma_{div} &=& \frac{1}{4\pi \epsilon}  \int d^d x\,
\sqrt{-g}\,
\left\{ \frac{24 +6N -n}{6} R  -
 g^{\mu\nu} \partial_\mu \phi \partial_\nu \phi [8-
n\Phi'(\phi)^2] \right. \nn \\
&& + \left.  {4\pi G \over \mu^\epsilon} e^{2\phi-f_2(\phi)}
(G^a_{\mu\nu})^2 [f_2'(\phi) -2 ] \right\},
\label{4}
\eea
where $\epsilon =d-2$.
It is convenient to introduce the following notations for the
divergent
 structures in (\ref{4}):
\beq
\Gamma_{div} =  \int d^d x\,  \sqrt{-g}\,
\left\{ R A_1 (\phi ) + g^{\mu\nu} \partial_\mu \phi \partial_\nu
\phi
 A_2 (\phi ) + {\mu^{-\epsilon} \over 4}
(G^a_{\mu\nu})^2 e^{-f_2(\phi)}  A_3 (\phi )  \right\},
\label{5}
\eeq
where the explicit form of $A_1$,  $A_2$ and  $A_3$ follows easily
from
(\ref{4}). Notice that  (\ref{4}) and (\ref{5}) in purely dilatonic
gravity
coincide with the $\Gamma_{div}$ obtained in the same gauge in
Refs.
\cite{8,2,5,7}. For the DYM theory,
$\Gamma_{div}$
(\ref{4}) coincides with the results of Ref. \cite{4}. The function
 $A_2 (\phi )$ in (\ref{5}) has been obtained in Ref. \cite{9},
too, but
this result is in disagreement with our result  (\ref{5}). This
could be the
indication of a mistake in \cite{9} (since that result is also
different from
the one obtained in \cite{8,2,4,5,7}) or could come from the use of
the
inconvenient background field method (in the
($2+\epsilon$)-dimensional,
non-covariant formalism). An explicit study of the gauge dependence
of
 $A_2 (\phi )$ in (\ref{5}) has been carried out in Refs.
\cite{8,10}.

Let us now investigate the RG properties of DYM
gravity
near two dimensions. By making use of the RG formalism in
$2+\epsilon$
dimensions \cite{9,10}, after some algebra one obtains the
following
$\beta$-functions
\bea
\beta_G &=& \epsilon G -\frac{1}{3} ( 48 +12N-2n) G^2, \nn \\
\beta_\Phi (\phi_0) &=& \mu {\partial \Phi (\phi_0)  \over
\partial \mu} = 8\pi \epsilon A_1 G (e^{2\phi_0}-1)\Phi'
(\phi_0) \nn \\
&&+ \frac{2\pi \epsilon^2 G}{\epsilon +1} [\Phi' (\phi_0) -1]
\int_0^{\phi_0} d\phi' \, e^{2\phi'}A_2(\phi'), \nn \\
\beta_{f_2} (\phi_0) &=& \mu {\partial f_2 (\phi_0)  \over
\partial \mu} = 8\pi \epsilon A_1 G (e^{2\phi_0}-1) f_2' (\phi_0)
\nn \\
&&+ \frac{2\pi \epsilon G}{\epsilon +1} [\epsilon f_2' (\phi_0)
+2(2-\epsilon)] \int_0^{\phi_0} d\phi' \, e^{2\phi'}A_2(\phi')-
\epsilon [A_3(\phi_0) -A_3(0)] , \label{6}
\eea
where $\phi_0$ is the bare dilaton field, and we are interested
in the study of the functional dependence of the $\beta$-functions
of the dilaton coupling functions. Observe that these beta functions are
of first order in $G$, which is used as a parameter in perturbation
theory.

As one can see rather easily, there exists a non-trivial
ultraviolet-stable fixed point at
\beq
G^* = \frac{3\epsilon}{48+12N-2n}, \ \ \ \  \epsilon > 0, \label{7}
\eeq
with the restriction $48+12N-2n >0$. Hence, by increasing the
number
of gauge fields, the scalar-matter central charge (i.e., the number
of
scalars) can be increased too. In this respect, let us remind the
reader that in pure dilatonic gravity with $n$ scalars one has
the restriction $n < 24$ \cite{9} and that in Einstein gravity
with $n$ scalars it is $n < 25$ \cite{3,14}. Hence, we obtain a
theory
of
quantum gravity with asymptotic freedom for the gravitational
constant
$G$. This theory can be considered as a quite nice toy model for
a consistent unification of gravity with matter. What is remarkable
is
the fact that the amount of matter in such a theory can be
significantly
augmented, as compared with all previous models of
($2+\epsilon$)-dimensional quantum gravity.

One can also study the fixed-point solutions of the beta functions
for
the dilatonic coupling functions. So, by using the following {\it
Ansatz} (motivated by string-model considerations)
\beq
\Phi (\phi) = \lambda \phi, \ \ \ \ \ \
f_2 (\phi) = \lambda_f \phi,
\label{8}
\eeq
and, substituting (\ref{8}) into  (\ref{6}), at the fixed point
(\ref{7})
one easily finds the following real solutions:
\beq
\lambda^* = -\frac{6\epsilon}{24+6N-n} + {\cal O} (\epsilon^2),
\ \ \ \  \lambda_f^* = -\frac{36\epsilon}{12+6N-n} +
{\cal O} (\epsilon^2).
\label{9}
\eeq
For the existence of a second fixed point in  (\ref{9}), in
addition to
the restriction $0<n<24+6N$ one has the condition $n\neq 12 +6N$.
A careful
study of the stability of the RG fixed point $(G^*, \lambda^* \phi,
\lambda_f^* \phi )$ shows that it corresponds to an ultraviolet
saddle point in the direction $\delta \Phi$ (see also \cite{10}).
\bs

\ni{\bf Higher-derivative dilatonic gravity near two dimensions}.
The next theory we want to discuss is HD dilatonic
gravity \cite{11}, a theory that is inspired by the study of the
massive modes in string theory. The corresponding action is
\bea
S& =& \int d^d x\,  \sqrt{-g}\,  \left\{ Z_1 (\phi )
 g^{\mu\nu} g^{\alpha\beta} \partial_\mu \phi
\partial_\nu \phi \partial_\alpha \phi
\partial_\beta \phi  + Z_2 (\phi )
 g^{\mu\nu} g^{\alpha\beta} \partial_\mu \phi
\partial_\alpha \phi \nabla_\nu \partial_\beta \phi \right. \nn \\
&& + Z_3 (\phi )
 g^{\mu\nu} g^{\alpha\beta} \nabla_\mu
\partial_\alpha \phi \nabla_\nu \partial_\beta \phi  + Z_4 (\phi )
 R g^{\mu\nu}  \partial_\mu \phi \partial_\nu \phi  + Z_5 (\phi )
 R^2 \nn \\
&& + \left. C_1 (\phi )  g^{\mu\nu} \partial_\mu \phi \partial_\nu
\phi
+ C_2 (\phi )  R - {1 \over 2} e^{-2\Phi (\phi )} g^{\mu\nu}
\partial_\mu
 \chi_{i} \partial_\nu \chi^i
 \right\}, \label{10}
\eea
where all dilatonic functions are assumed to be smooth enough.
In order to make the theory multiplicatively renormalizable
in the generalized sense in two
dimensions \cite{11}, the initial action should contain a term of the
form
$Z_6 (\phi ) R \Delta \phi$. However, we shall restrict our
analysis here
to one-loop order and one can show that the theory (\ref{10}) is
one-loop renormalizable even without the above term, so that we are
allowed to drop it for simplicity. Moreover, the kinetic term for
the
dilaton may be always eliminated by a conformal transformation of
the
metric, and hence we can set above $C_1 (\phi ) =0$ as well ---as
in
the preceding case of the DYM  theory.

Using the background-field method and working in the conformal
gauge,
we can split the fields as
\beq
\phi \longrightarrow \phi  + \varphi, \ \ \
\chi_i \longrightarrow \chi_i  + \eta_i, \ \ \
g_{\mu\nu} \longrightarrow g_{\mu\nu}  \, e^\sigma,
 \label{11}
\eeq
where $\varphi$, $\eta_i$ and $\sigma$ are now the quantum fields.

The calculation of the one-loop effective action can be performed
in close analogy with Ref. \cite{11}. One has to obtain the Tr log
of the operator, which appears after expanding (\ref{10}), over the
quantum fields near the background  (\ref{11}), keeping up to
second-order terms:
\beq
\wh{H}_{ij} = \wh{K}_{ij} \Delta^2 + \wh{L}_{ij}^\lambda \Delta
\nabla_\lambda + \wh{M}^{\mu\nu}_{ij} \nabla_\mu \nabla_\nu +
\cdots
\label{12}
\eeq
the action of these operators being on the space $\left\{ \varphi
;
\sigma; \eta_i \right\}$. Power-counting arguments show that
low-order terms in (\ref{12}) are irrelevant for the calculation of
the one-loop divergences. Owing to the fact that the matrices
$\wh{K}$,  $\wh{L}$ and $\wh{M}$ in  (\ref{10}) do not commute,
generally
speaking, with the covariant derivatives, integration
by parts can change the naive choice of their elements. The
standard
method in order to make of the operator (\ref{12}) a unique
Hermitian
operator is to use the doubling procedure of 't Hooft and Veltman
\cite{12}.
Another problem is caused by the fact that the matter fields only
contribute
to the second derivatives in  (\ref{12}) and, therefore, $\wh{K}$
is
degenerate. But this can be improved by means of operator squareing
\beq
\wh{H} \longrightarrow \wh{\cal H} = \wh{H} \wh{\Omega}, \ \ \ \ \
\
 \wh{\Omega} = \left( \brr{ccc} 1 & 0 & 0 \\ 0 & 1& 0 \\
0 & 0 & \delta_{ij} \Delta \err \right), \label{13}
\eeq
with the subsequent subtraction of the squareing operator
$\wh{\Omega}$.
Taking all this into account, the actual calculation is done along
the
same lines as in  Ref. \cite{11} (and, hence, we shall skip all
details).

Let us just briefly recall the different steps involved. The Tr log
of
the operator $\wh{\cal H}$ can be found with the use of the
algorithm (see, for example, \cite{8})
\bea
&& i \tr \log \left( \wh{1} \Delta^2 + \wh{E}^\lambda \Delta
\nabla_\lambda
+\wh{\Pi}^{\mu\nu} \nabla_\mu \nabla_\nu + \cdots \right)_{div} \nn \\
&& =-\frac{1}{4\pi \epsilon}  \int d^2 x\,  \sqrt{-g}\,  \left[
\frac{1}{4} \tr \left( \wh{E}^\lambda \wh{E}_\lambda\right) -
\frac{1}{4} \tr \wh{\Pi}_\nu^\nu - \frac{R}{3} \tr \wh{1} \right],
\label{14}
\eea
modulo non-essential surface terms.

Calculating (\ref{14}) and extracting from it the contribution of
the squareing operator and the standard Polyakov ghost
contribution in the conformal gauge, one obtains finally
\bea
\Gamma_{div} &=& \frac{1}{4\pi \epsilon} \int d^2 x\,  \sqrt{-g}\,
\left[ \left( \frac{7-n}{6} + \frac{Z_4}{Z_3} \right) R + \left(
n(\Phi')^2 + \frac{3Z_3}{8Z_5} + \frac{4Z_1}{Z_3}- \frac{Z_4}{2Z_5}
- \frac{Z_4^2}{2Z_3Z_5} \right. \right. \nn \\
&&- \left. \left. \frac{Z_2'}{Z_3} + \frac{Z_3''}{Z_3}-
\frac{Z_2Z_3'}{2Z_3^2} - \frac{(Z_3')^2}{2Z_3^2} \right)
  g^{\mu\nu} \partial_\mu \phi \partial_\nu \phi \right] \nn \\
&\equiv&  \int d^2 x\,  \sqrt{-g}\, \left[ A_1(\phi) R  +A_2(\phi)
  g^{\mu\nu} \partial_\mu \phi \partial_\nu \phi \right].
\label{15}
\eea
One thus sees that the theory (\ref{10}) is one-loop renormalizable
and also that for some choices of the dilatonic functions $Z_i$ it
is one-loop multiplicatively renormalizable, or even one-loop
finite. The fact that the theory (\ref{1}) is not conformally
invariant at the classical level leads to a non-standard structure
of (\ref{15}). In particular, the renormalization of the curvature
in (\ref{15}) becomes dilatonic-couplings dependent and hence the
analog of the conformal anomaly coefficient is dilatonic dependent,
and, presumably, it can be interpreted as a central charge only at
the conformal fixed point. This is seen also from the fact that
naive counting of the central charge (the coefficient of $R$)
---which works perfectly for standard dilatonic gravity of type
(\ref{1})--- is no longer correct for a theory as (\ref{10}).

One can, nevertheless, try to find the regimes for which theory
\ref{10}) approaches the conformally invariant regime at the
one-loop  level (being, as it is, multiplicatively renormalizable).
This gets support from the fact that we know of a subclass of the
family (\ref{10}) which is conformally invariant. It is namely the
theory  with $Z_5=$ const. and all the remaining $Z_i=0$,  without
a low-derivative dilatonic sector. This theory can be mapped into
a particular form of low-derivative dilatonic gravity (see
\cite{16}).

Choosing the HD dilatonic couplings to be constant,
$Z_i(\phi)=Z_i=$ const., and imposing the finiteness conditions, one
obtains:
\beq
 \frac{7-n}{6} + \frac{Z_4}{Z_3}=0, \ \ \ \
n[\Phi'(\phi) ]^2 + \frac{3Z_3}{8Z_5} + \frac{4Z_1}{Z_3}-
\frac{Z_4}{2Z_5} - \frac{Z_4^2}{2Z_3Z_5} =0,
\label{16}
\eeq
where $\Phi'(\phi)$ should be also choosen to be constant for
consistency. As one can see, there are in general many possible
solutions satisfying the finiteness conditions (\ref{16}) (we have
just two equations and five free parameters). In the regime
(\ref{16}) our theory goes at one-loop, presumably, to a
conformally invariant phase of the theory (as in non-linear sigma
models).

Let us now consider the theory (\ref{10}) in $2+\epsilon$
dimensions, imposing on the functions $Z_i$ the only restriction
that $Z_4(\phi) /Z_3 (\phi) = Z =$ const. and choose $C_1(\phi)=0$,
$C_2(\phi)=\frac{\mu^\epsilon}{16\pi G} e^{-2\phi}$. Then, using
again the fact that the theory (\ref{10}) has a smooth $\epsilon
\rightarrow 0$ limit (as was the case with (\ref{1})), one can see
that the one-loop effective action (\ref{15}) corresponds also to
the one-loop effective action near two dimensions. Since the HD
sector is not renormalized at all at the one-loop level, we may
restrict our discussion of the RG in $2+\epsilon$ dimensions to the
low-derivative sector. (Of course, this should be significantly
modified at two-loop order, where a non-trivial renormalization of
the HD dilatonic function couplings is expected.) Explicit
calculation shows that the RG $\beta$-functions are given by
(\ref{6}) with the only change of $A_1$ and $A_2$ in accordance
with (\ref{15}). Then, one can see that, again, there is a
non-trivial, ultraviolet-stable fixed point for the gravitational
coupling constant
\beq
G^* = \frac{3\epsilon}{14-2n+6 Z_4Z_3^{-1}}, \ \ \epsilon >0,
\label{17}
\eeq
where $14-2n+6 Z_4Z_3^{-1}>0$. Now, the asymptotic freedom of the
gravitational coupling constant can be achieved for any number of
scalars by choosing conveniently the ratio of the two dilatonic
couplings that appear in (\ref{17}).

One can also look for the existence of a non-trivial fixed point
for the scalar-dilaton coupling in (\ref{10}). Choosing for $Z_i
(\phi)$ and for $\Phi (\phi)$ an {\it Ansatz} similar to (\ref{8}),
i.e.,
\beq
\Phi (\phi) = \lambda \phi, \ \ \ \ Z_i(\phi ) = e^{\lambda_Z
\phi}, \ i \neq 4, \ \ \ \ Z_4 (\phi ) = Z  e^{\lambda_Z \phi}, \
Z = \mbox{const.},
\label{18}
\eeq
and substituting it into the $\beta$-function for $\Phi$, we can
look for fixed-point solutions.

Owing to the fact that $\lambda_Z$ and $Z$ are free parameters at
one-loop level, we immediately find a solution of the same
nature as (\ref{8}):
\beq
\lambda^* = \frac{3\epsilon}{8} \, \frac{\frac{19}{4}-Z-Z^2+\lambda_Z^2
- 3\lambda_Z}{7-n+6Z},
\label{19}
\eeq
under the hypothesis that $Z$ and $\lambda_Z$ are ${\cal O} (1)$
quantities. Hence, we see that the lower-derivative sector of the
theory (\ref{15}) possesses a non-trivial ultraviolet fixed point
at which asymptotic freedom of the gravitational coupling constant
is realized. These nice properties of the theory are a consequence of
the huge freedom that exists in choosing the parameters (e.g. the
dilatonic couplings). Surely, this choice will become more
constrained at the two-loop level. Furthermore, the gravitational
coupling constant is no more the natural parameter of perturbation
theory (unlike in DYM  gravity) and so the
asymptotic freedom on $G$ alone is not sufficient for the good
ultraviolet behavior of the theory. Finally, one should point out
that the possibility of asymptotic freedom for $G$ may be
destroyed at the multi-loop level, where other connections between
$Z_3$ and $Z_4$ are to be expected from the renormalization of these
functions.
\bs

\ni{\bf Conclusions}.
In summary, we have shown that both DYM  gravity and
HD dilatonic gravity near two dimensions exhibit a non-trivial,
ultraviolet fixed point with an asymptotically free gravitational
coupling constant. The behavior of the dilatonic couplings at the
fixed point is defined. The addition of the Yang-Mills sector to
dilatonic gravity seems to be more promising ---with the purpose of
increasing the matter central charge--- than the introduction of the
HD terms, due to the fact that the nice properties of HD dilatonic
gravity may be completely destroyed at the multi-loop level (what
is certainly not the case for DYM gravity). Therefore, it would be of
interest to study the model (\ref{1}) beyond two dimensions using
dynamical triangulations in higher dimensions (see, for
example, \cite{17}).

Notice, finally, that by a conformal tranformation of the metric,
$g_{\mu\nu} \rightarrow g_{\mu\nu} \exp \left(
\frac{4\lambda^*}{\epsilon} \phi \right)$, the theory (\ref{1}) can
be presented at the fixed point under the form of a charged
string-inspired model with the scalars being non-interacting with the
dilaton (the CGHS model \cite{15} with a Yang-Mills sector). On the
other hand, by means of the transformation $g_{\mu\nu} \rightarrow
g_{\mu\nu} \exp \left( \frac{2\epsilon \lambda^*_f}{2-\epsilon}
\phi \right)$, at the non-trivial fixed point the theory (\ref{1})
acquires the form of a theory where the Yang-Mills sector does not
interact with the dilaton (however, a kinetic term for the dilaton
appears).
\vspace{5mm}


\noindent{\large \bf Acknowledgments}

We  are indebted with S. Naftulin for very useful comments on the
subject of HD dilatonic gravity, and with J.I. Latorre, N. Sakai,
Y. Tanii and A. Tseytlin for interesting discussions.
This work has been supported by DGICYT (Spain), project Nos.
PB93-0035
and SAB93-0024, and by CIRIT (Generalitat de Catalunya).






\end{document}